\documentclass[%
prb,
 jmp,%
 amsmath,amssymb,
 reprint,%
]{revtex4-1}

\usepackage{graphicx}
\usepackage{dcolumn}
\usepackage{bm}
\usepackage{setspace}

\begin{document}

\preprint{AIP/123-QED}

\title{Testing The Existence of Single Photons}

\author{Quynh M. Nguyen}


\author{Asad Khan}

\affiliation{University of Minnesota, Minneapolis, MN, USA, 55455}

\date{May 12, 2014}

\begin{abstract}
We demonstrated the existence of single photon by counting correlation of laser output of a birefringent $\beta$-Barium Borate (BBO) crystal. The best result of anti-correlation parameter violates the classical prediction for wavelike behavior of light by 80 standard deviations. 
\end{abstract}

\maketitle

%

\section{\label{sec:level1}Introduction}
Classically, light is an electromagnetic wave. Quantum mechanically, light has both wave-like and particle-like properties, and the particle-like quanta are called photons. Experiments such as the photo-electric effect can make this particle-like nature of the field apparent but that is still not enough\cite{Stanley} to prove that light is made of photons\cite{Thorn}. The reason is, we never measure light directly but always measure the associated photocurrent instead. It is possible for the granularity appearance to be caused by the discrete nature of the electron in the detectors. Thus, observing granularity in measurements of the field is necessary but not a sufficient proof for the existence of photon.

In 1986, Grangier, Roger, and Aspect\cite{Grangier} performed an experiment in which they examined correlations between photodetections at the transmission and reflection outputs of a 50/50 beamsplitter. If a single quantum of light is incident on the beamsplitter (BS), it should be detected at the transmission output or at the reflection output, but not both: there should be no coincident detections between the two outputs. In fact, Grangier et al. measured fewer coincidences than predicted by a classical wave theory. Their result violates a classical inequality by 13 standard deviations, demonstrating that the field incident on the beamsplitter was well described by a single-photon state. The key difficulty in such a measurement is to create a field that truly has a single-photon incident on the BS. A dim source of light containing \textit{on average} a single photon is not sufficient.

Since photons are quantum mechanical objects, an experiment which requires a quantum mechanical explanation would imply that there is more to the field than just classical wave. If classical theories of electromagnetism cannot explain the result, we take it to mean that single photons exist.

Our experiment is an updated version of the one performed originally by Grangier et al., using a higher count rate and more advanced equipment. The increase count rate allows a violation of classical inequality by 80 standard deviations with 10 minutes of counting time. The experiment is well described by the quantum mechanics of a field in a single photon state incident upon a BS.
\begin{figure}
\includegraphics[scale=0.3]{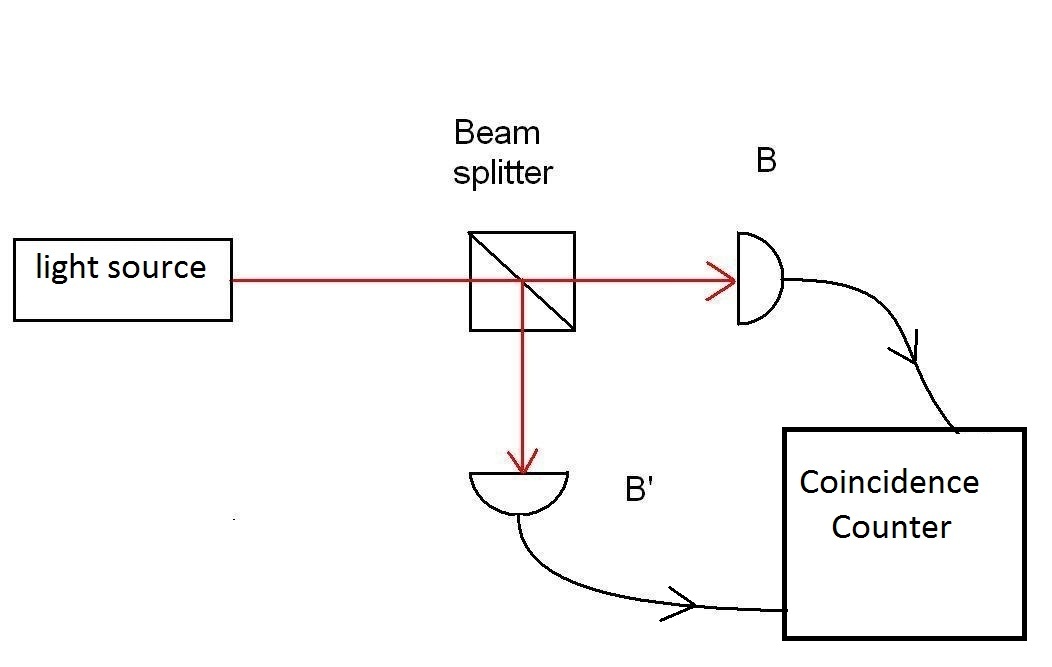} 
\caption{Coincidence measurement. The incident beam is split into transmitted (detected at B) and reflected (detected at B$'$) at a 50/50 beamsplitter (BS). Detection at B and B$'$ are examined to se whether or not they occur simultaneously}
\label{history}
\end{figure}
\section{Theory}
\subsection{Early measurements}
We are interested in examining correlations between the photocounts on two detectors situated at the output ports of a BS (Fig. \ref{history}). The first experiment to examine these correlations was carried out by Hanbury Brown and Twiss \cite{Hanbury} who found a positive correlation between the outputs of the two detectors. However in their experiment, Hanbury et al. were not counting individual photons, but were instead measuring correlations between continuous currents from photomultiplier tubes (PMTs) \cite{Hanbury}. This positive correlation only indicated that when the current from one PMT increased, the current on the second tended to increase simultaneously. 

In contrast, when Brannen and Ferguson performed a similar experiment, they observed no positive correlation, and then claimed ``that if such a positive correlation did exist, it would call for a major revision of some fundamental concepts in quantum mechanics."\cite{Brannen} However, Hanbury Brown and Twiss\cite{Hanbury2} noted that the experimental parameters used by Brannen and Ferguson precluded the observation of positive correlations.  They showed that positive correlations are not only allowed by quantum mechanics, but also are a natural consequence of the tendency for photons and other bosons to ‘‘bunch’’ together. 

The first experiment to observe positive correlations using coincidence detection of individual photocounts (not just photocurrents) from PMTs was performed by Twiss, Little, and Hanbury Brown \cite{Hanbury3}. They observed positive correlations of a few percent. This amount of correlation was expected, given their experimental parameters.
\subsection{\label{sec:level2}Classical field}
By classical field, we mean an electromagnetic wave that is completely described by Maxwell’s equations. Referring to Fig. \ref{history}, the intensity of the field incident on the polarizing beamsplitter (PBS) is I(t), the transmitted and reflected fields from the PBS go to detectors B and B$'$, and their intensities are $I_B(t)$ and $I_{B'}(t)$. The correlation between $I_B(t)$ and $I_{B'}(t)$ are given by anti-correlation parameter\footnote{as known as degree of second-order coherence} $\alpha^{2d}(\tau)$, which is a function of the time delay $\tau$ between the intensity measurements:
\begin{equation}
\alpha^{2d}(\tau) = \frac{\langle I_B(t+\tau)I_{B'}(t) \rangle }{\langle I_B(t+\tau)\rangle \langle I_{B'}(t) \rangle} 
\label{eq:intensity}
\end{equation}
The superscript 2d indicates a two-detectors set up. If the light source is stationary, then the brackets are interpreted as ensemble averages rather than time averages. We are interested in simultaneous ($\tau$ = 0) intensity correlations. If the beamsplitter has an intensity transmission of Tr, and reflection coefficient of Re, then the transmitted, reflected, and incident intensities are related by
\begin{equation}
I_B(t) =Tr \times I(t)   \text{ and }    I_{B'} = Re \times I(t)
\end{equation}
Inserting these expressions into eq. (\ref{eq:intensity}) gives
\begin{equation}
\alpha^{2d}(0)= \frac{\langle [I(t)]^2 \rangle }{ \langle I(t) \rangle ^2}
\end{equation}
which is independent of the splitting ratio of the beamsplitter. The variance of the intensity, which can be written as $\Delta I^2 = \langle I^2 \rangle-\langle I \rangle^2$, must be a positive number\cite{Loudon}. Thus, it must be true that
\begin{equation}
\alpha^{2d}(0)\geq 1 \text{ (classical fields)}\label{eq:classical}
\end{equation}
This result has been derived using classical wave theory. In eq. (\ref{eq:classical}), equality with 1 is achieved if the input intensity is perfectly stable with no fluctuations.
The closet experimental realization of a stable, classical wave is laser light. Light from a stable laser yields $\alpha = 1$, no matter how dim it is.
\subsection{\label{sec:level2}Semi-classical theory of photodetection}
In the previous section, anti-correlation is based on intensities of the fields leaving the beamsplitter. However, in practice, one does not measure intensities directly but rather the photocurrent from detectors. Therefore, it is neccesary to model the photodetection process. In semi-classical theory of photoelectric detection, the field is treated classically and the photodetector is treated quantum mechanically\cite{Mandel}. Accordingly, the conversion of continuous electromagnetic waves into discrete photoelectrons is a random process. The probability of obtaining a single photocount from a single detector (for example, detector B), within a short time window $\Delta t$, is proportional to the average intensity of the incident field:
\begin{equation}
P_B = \eta_B \langle I_B(t)\rangle \Delta t \label{eq:B prob}
\end{equation}
where $\eta_B$ is a constant that characterizes the the detection efficiency of detector B. The joint probability of getting a photocount at detector B’ (within a short time window $\Delta t$), and then after a time $\tau$ obtaining a photocount at detector B (also within the time window $\Delta t$) is given by
\begin{equation}
P_{BB'} = \eta_B\eta_{B'} \langle I_B(t+\tau) I_{B'}(t)\rangle \Delta t \label{eq:B' prob}
\end{equation}
Combining eqs. (\ref{eq:intensity}), (\ref{eq:classical}), (\ref{eq:B prob}), and (\ref{eq:B' prob}) gives
\begin{equation}
\alpha^{2d}(0)= \frac{P_{BB'}(0)}{P_B P_B'} \geq 1 \text{ (classical fields) } \label{eq:classic 2}
\end{equation}
In summary, it is possible to measure anti-correlation parameter between the fields leaving a BS by measuring the probability of joint and individual photocounts at detectors B and B$'$. And the anti-correlation parameter must satisfy $\alpha^{2d}(0) \geq 1$. The probability and $\alpha^{2d}(0)$ are modeled after our experiment set up. \section{\label{sec:level1}Experimental technique}
\subsection{The experimental probabilities}
Experimentally, probabilities must be expressed in terms of measured count rates. For examples, the probability of a detection at B in a short time interval $\Delta t$ (characteristic of the hardware in use) is simply given by the average rate of detections multiplied by $\Delta t$. The average rate of detections at B is just the number of detection $N_B$ divided by the total counting time T. The probabilities for B’ detections and BB’ coincidences are given similarly:

\begin{equation} 
P_B =    \frac{N_B}{T}  \Delta t \text{, }P_{B'} =    \frac{N_{B'}}{T}  \Delta t \text{, } P_{BB'} =    \frac{N_{BB'}}{T}  \Delta t \label{eq:exp prob} 
\end{equation}

Here $N_{BB'}$ is the number of coincidence counts between the two detectors. Substituting eq. (\ref{eq:exp prob}) into eq. (\ref{eq:classic 2}) gives
\begin{equation}
\alpha^{2d}(t)  =  \frac{N_{BB'}}{N_B N_{B'}} \times \frac{T}{\Delta t} \text{  (2-detectors)}
\end{equation}
Thus, with classical field statistics, it is possible to measure anti-correlation between the fields leaving PBS by measuring the probability of joint and individual photoncounts at detector B and B$'$. The time window $\Delta t$ and the total measuring time T are adjustable. Alpha must satisfy the inequality $\alpha \geq 1$.  When the photocounts at B and B$'$ are completely uncorrelated $\alpha = 1$, which occurs when the input field to the PBS is perfectly stable. When the input field fluctuates, then $\alpha > 1$ indicating positive correlations. 

\subsection{Experimental apparatus}
\begin{figure}
\includegraphics[scale=0.29]{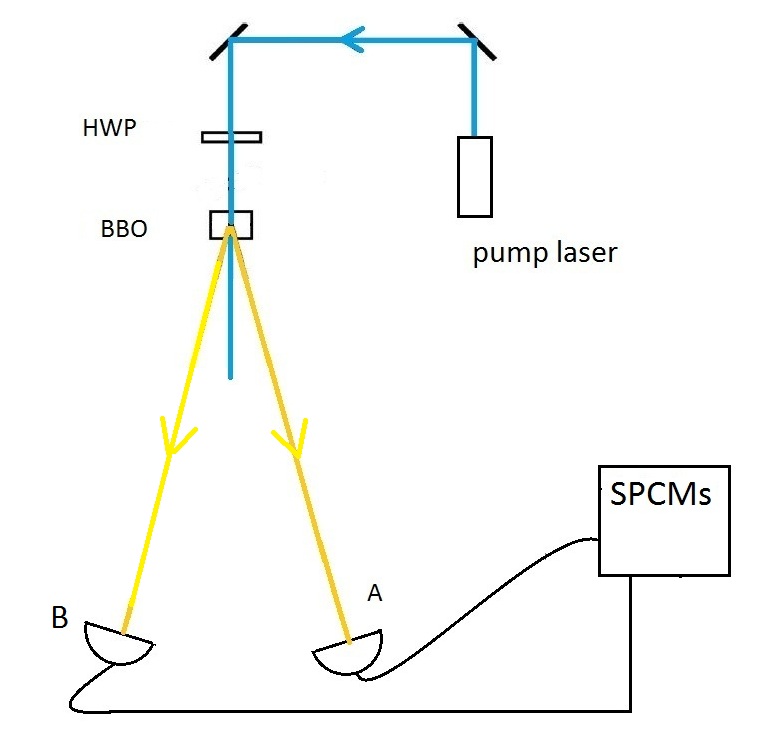}
\caption{The set up for anti-correlation measurement of parametric spontaneous downconverted lights. Major components include 405nm pump laser (blue), Downconversion Crystal (BBO), single photon counting modules (SPCMs), fiber-coupling lens A, B and half wave plate (HWP). Optical fibers direct the light from A,B to their corresponding SPCMs. Downconverted lights make a $3^o$ angle with respect to the main beam and are shown in yellow. The distance from the BBO to detector A is 100cm. The correlation between photons dectected at detector A and B must be established to ensure single photon incidence on the PBS.}
\label{2d}
\end{figure}

\begin{figure}
\includegraphics[scale=0.3]{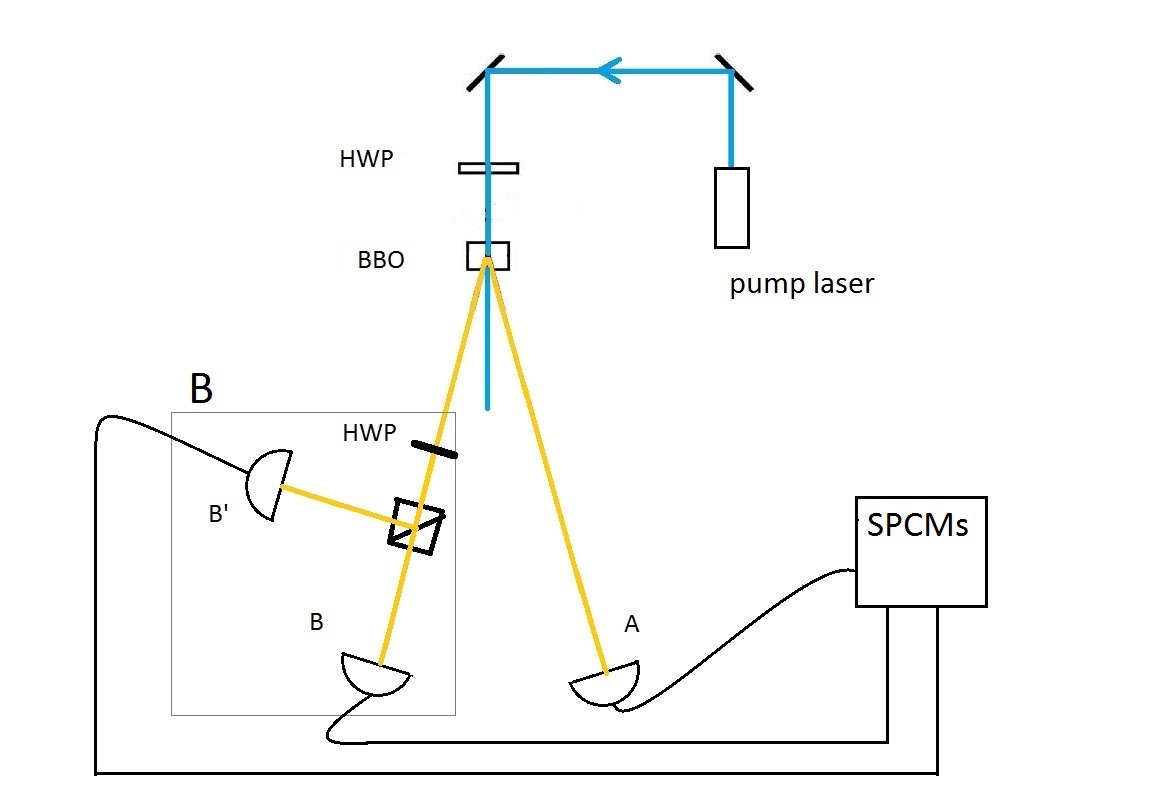} 
\caption{Three-detectors experimental apparatus. A polarized beamsplitter (PBS) and detector B$'$ were added to the two-detectors apparatus.}
\label{single_f}
\end{figure}

In order to measure $\alpha = 0$, it is necessary to have a single photon incident on the PBS shown in Fig. \ref{history}. We exploited spontaneous parametric downconversion process to produce single-photon states. The experiment use a $\beta$-barium borate (BBO) crystal cut at $29^o$ so that the twin photons emerge with approximately equal energy ($\lambda$ = 810nm) at $3^o$ from the initial pump direction \cite{Pearson}. Anti-correlation parameter was measured for the two down-converted beams using detector A and B (Fig. \ref{2d}). These two beams need to be higly correlated in order to use photodetections at A as a condition for single photon incident on the PBS.
In the final set up (shown in Fig. \ref{single_f}), a single photon incident on the PBS is achieved conditionally upon the detection of a photon at detector A. Thus, three-detectors measurements are necessary to see $\alpha =0$. Using the same source without conditioning, detectors B and B$'$ would detects non-single photon fields and give $\alpha \geq 1 $. In a real experiment, $\alpha$ is expected to be non-zero because there will always be some accidental three-fold coincidence counts $N_{ABB'}$. Next, $\alpha$ needs to be modeled after a three-detectors set up. 
\subsection{Three-detectors anti-correlation}
If the measurement of $\alpha^{2d}$ is conditioned on a measurement of photocount on a third detector A, in place of eq. (\ref{eq:classic 2}), we have 
\begin{equation}
\alpha^{3d}(0)  =  \frac{P_{ABB'}(0)}{P_{AB}(0) P_{AB'}(0)} \text{ (3-detectors)}
\end{equation}
Where $P_{ABB'}$ is the probability of threefold coincidence detection, and $P_{AB}(0)$ and $P_{AB'}(0)$ are the probabilities of coincidence detection between detector A and detectors B and $B'$, respectively. The superscript 3d indicates three-detector anti-correlation parameter.
Since the events of interest are conditioned on events at A, the number of detections at A, $N_A$, serves as the total number of trials, which can be used to normalized probabilities :
\begin{equation}
P_{AB}(0) =    \frac{N_{AB}}{N_A}   \text{, } P_{AB'}(0) =    \frac{N_{AB'}}{N_A}  \text{, } P_{ABB'}(0) =    \frac{N_{ABB'}}{N_A}  
\end{equation}
Using these probabilities, $\alpha^{3d}$ becomes
\begin{equation}
\alpha^{3d}(0)  =  \frac{N_{ABB'}}{N_{AB} N_{AB'}}N_A \text{  (3-detectors)} \label{eq:last 3d}
\end{equation}
For classical fields, $\alpha^{3d} \geq 1$. If a single quantum of light is incident on the PBS, it should be detected at the transmission output or at the reflection output, but not both: one photon should not be detected twice. In other words, $N_{ABB'} = 0$ in eq. (\ref{eq:last 3d}). Quantum mechanics predicts $\alpha = 0 $ for 3 detectors.
\begin{table}[h]
\centering
\setlength{\tabcolsep}{7.5pt}
\begin{tabular}{l*{4}{c}r}
 \hline
\hline
$\Delta t$ & $R_{A}$ & $R_{B}$  & $R_{AB}$ & $R_{AB}^{acc}$  & $\alpha^{2d}$ \\
(ns) & (Hz)  & (Hz) & (Hz)& (Hz) &  \\ 
\hline
10 & 14,800 & 16,700 & 223 & 2.48 & $90.2\pm4$ \\
20 & 14,600 & 16,700 & 231 & 4.77 & $48.5\pm0.6$\\
40 & 14,200 & 15,700 & 238 & 8.99 & $26.5\pm0.4$ \\
60 & 14,100 & 15,700 & 244 & 13.4 & $18.3\pm0.9$\\
\hline 
10 & 46,100 & 51,300 & 719 & 23.7 & $17.9\pm0.2$ \\
20 & 45,600 & 50,700 & 756 & 46.2 & $16.4\pm0.2$\\
40 & 44,300 & 49,200 & 803 & 87.1 & $9.2\pm0.1$ \\
60 & 43,800 & 48,600 & 843 & 128 & $6.6\pm0.1$\\

\hline
\hline
\end{tabular} 

\caption{Correlation results for the two arms of our down-converted light sources at two different attenuation levels. We report averages of 20 runs, each 30 s long. The values of of the anti-correlation parameter indicate that the two arms are correlated.}
\label{table 2d}
\end{table}

\begin{table*}[t]
\centering
\setlength{\tabcolsep}{12pt}
\begin{tabular}{l*{6}{c}r}
\hline
\hline
$\Delta t$    & $R_A$ & $R_{AB}$ & $R_{AB'}$ & $R_{ABB'}$ & $R_{ABB'}^{acc}$  & $\alpha^{3d}$ & violation \\
 (ns) &(Hz)  & (Hz) & (Hz)& (Hz) & (Hz)  & & $(\sigma)$ \\
\hline
10 & 14,800 & 128 & 95 & 0.016 & 0.018 & $0.019\pm0.012$ & 80  \\
20 & 14,600 & 132 & 99 & 0.031 &  0.037 & $0.035\pm0.015$ & 62  \\
40 & 14,200 & 136 & 102 & 0.062 &  0.075 & $0.064\pm0.018$ & 53  \\
60 & 14,100 & 140 & 104 & 0.10 &  0.11 & $0.097\pm0.023$ &  40  \\
\hline
10 & 46,100 & 412 & 306 & 0.19 & 0.18 & $0.71\pm0.02$ & 47 \\
20 & 45,600 & 431 & 325 & 0.34 & 0.38 & $0.112\pm0.04$ & 22\\
40 & 44,300 & 458 & 345 & 0.76 & 0.78 & $0.213\pm0.04$ & 18\\
60 & 43,800 & 477 & 366 & 1.1 & 1.2 & $0.268\pm0.33$ & 22 \\
\hline
\hline
\end{tabular} 
\caption{Correlation results for a three-detectors measurement using a down-converted light source (all rates measure in Hz) at two attenuation levels. We report averages of 20 runs, each 30 s long. Last column shows violation of the classical inequality $\alpha^{3d} \geq1$ in unit of $\sigma$}
\label{table 3d}
\end{table*}
\subsection{Accidental coincidence}
In principle, our detectors can measure single photons. However, all detectors and their associated electronics have physical limitations such as efficiency and resolution that can affect the experimental result. Single-photon detectors convert an incoming photon to an electronic pulse. The pulse widths $\tau_p$ used were in order of 10 ns. For the counting circuit to register a coincidence, the pulses from two detectors must overlap at least partially. If the pulse widths are the same, the time window is $\Delta t = 2 \tau_p$.
For two random and independent sources, all coincidence counts are purely accidental. If the average count rates for detector A and B are $R_A$ and $R_B$, the expected rate of accidental coincidences is given by\cite{Eckart}
\begin{equation} \label{R}
R^{acc}(2d)=\Delta t R_A R_B \text{ (2 detectors)}
\end{equation}
With a three-detectors set up, one possibility of accidental coincidence is for three uncorrelated photons to end up at detector A, B, B$'$ within a small enough time such that the pulses overlap. However, for small pulse widths, such threefold accidentals are extremely rare because it depends on the square of $\tau_p: R_{acc}= 3\tau_p^2 R_AR_B R_B'$.\cite{Eckart}. 
Much more likely is a twofold random accidental between a real twofold coincidence and a random third single. With background negligible (by using filters), the threefold accidental rate can be estimated by
\begin{equation} \label{R3}
R^{acc}(3d) \approx \Delta t R_A R_B \text{ (3 detectors)}
\end{equation}
Using eq. (\ref{R}) and eq. (\ref{R3}), the rate of accidental coincidences were estimated and included in calculating $\alpha$.
\section{\label{sec:level1}Result}
A correlation measurement using two detectors A and B corresponding to both arms of down-converted light was performed. Because the beams are correlated, it was expected that number of coincidence to be greater than those of random source, that is, $\alpha^{2d} > 1$. Typical results from 20 30s-long runs are summarized in Table \ref{table 2d}. Data in Table \ref{table 2d} clearly shows that $\alpha^{2d}$ is larger than one, indicating the down-converted light is correlated and appropriate for producing single-photon state.

Typical results for the three-detector correlation experiment using the verified single photon source are shown in Table \ref{table 3d}. Values of $\alpha^{3d}$ were found to be less than one, which provides clear evidence for the quantum nature of light. $\alpha^{3d} <1 $ means light consists of quanta that can be transmitted or reflected at a beam splitter, but not both. 
\section{\label{anal}Analysis}
\subsection{Down-conversion correlation}
In theory, down-converted lights incident on detector A and B are perfectly correlated, and $\alpha^{2d}$ should be much larger than the result shown in Table \ref{table 2d}. Consider an ideal experiment with perfect alignment and detectors and no background noise. In this case, every twin pair produced will be measured and all pairs are in perfect coincidence. If the rate of twins incident on the detector is R(t), then $R_A=R_B=R_{AB}=R(t)$. Using eq. (\ref{2d prob}) and the fact that the rate of event is just the total number of count divided by the total integration time T: $R_{i}=\frac{N_{i}}{T}$, $\alpha^{2d}$ for the ideal experiment is given by
\begin{equation} \label{2d acc}
\alpha^{2d} = \frac{R_{AB}}{R_A R_B \Delta t} = \frac{1}{\Delta t R(t)}
\end{equation}
For a rate of 14,800 Hz and $\Delta t =$ 10ns, perfect equipment would give $\alpha^{2d}=6757$. The efficiency of this kind of system is about 4\% \cite{Pearson}, leaving  $\alpha^{2d}$ at most ~270, in the same order of magnitude with our measurement. 

Interestingly, $\alpha^{2d}$ can be increased by decreasing the rate of down-converted lights, as can be seen in Table \ref{table 2d}. The reason is that the accidental coincidence rate depends quadratically on R(t) in eq. (\ref{R}), while the measured coincidence rates depends linearly on R(t) in eq. (\ref{2d acc}).
\subsection{Anti-correlation parameter with three detectors}
If a truly single-photon state were incident on the PBS, quantum mechanics would predict that $\alpha^{3d} = 0$. A consequence of defining a ``coincidence" within a finite time window $\Delta t$ is an expected nonzero anti-correlation parameter. This is because the possibility that uncorrelated photon from different downconversion events may hit the B and B$'$ within the time window. These are accidental coincidences, which increase with the time window. Data in Table \ref{table 2d} ad Table \ref{table 3d} show that as the count rate and time window increase, so do the rate of accidental coincidence $R_{AB}^{acc}$ and $R_{ABB'}^{acc}$.

The dependence on time window of accidental coincidence also explains why the conindicence count $R_{ABB'}$ decreases as the time window $\Delta t$ decreases. A perfectly correlated twin coincidence should not depend on $\Delta t$, but in this case, a portion of the experimental coincidences is accidental.

To improve correlation measurement, accidental coincidence must be reduced. One method is to reduce the coincidence window. This is supported by the date in Table \ref{table 3d}, but it is limited by hardware. A second method is to reduce the power of the pump laser, which effectively lowers all count rates in eq. (\ref{R3}).

\subsection{Other considerations}
Our best value of $\alpha^{2d}$ for the downconverted beams is about three times smaller than theoretical prediction (90 compare to 270). The efficiency of our system is lower than that of Pearson et al.,\cite{Pearson} who used similar equipment. Better precision of measuring $\alpha$ can be made by improving alignment technique. Especially, we found difficulty in defining the beam path (a necessary step for positioning the PBS) of a non-visible laser using irises.

Table \ref{table 2d} also shows that $R_A$ and $R_B$ are not equal on average. The problem pertained after exchanging the fiber-coupling lens. Thus, the SPCM corresponds to detector B must have a higher efficiency than that of detector A. Coincidence rate can be increased if detector A is improved.

Given the beam path's length (100cm from the BBO to fiber-coupling lens A), the time delay effect due to uncertainty in path's length (2mm) is $\approx 0.01$ns, which is negligible compared to our smallest time window (10ns).
\section{\label{con}conclusion}
We have performed an experiment whose results cannot be explained using classical theory of electromagnetic wave. The results are consistent with a quantum mechanical description of light in which a field in a single-photon state is incident on a beamsplitter. Therefore, we take the result as a verification of the existence of photons. 
\section{\label{ack}acknowledgement}
We are grateful to Kurt Wick for his support and guidance. We also would like to thank professor G. Pawloski and professor C. Pryke for their reviews of our Technical Design Report.

\end{document}